\documentclass[a4paper]{article}

\usepackage{icrc2013}

\title{Studying flux variability of the BL Lac object 1ES0806+524 with MAGIC in a multi-wavelength context}

\shorttitle{Spectral variability studies of the BL Lac object 1ES0806+524 with MAGIC}

\authors{
K. Berger$^{1}$,
C. Schultz$^{2}$,
E. Lindfors$^{3,4}$
R. Rheinthal$^{4}$,
A. Stamerra$^{5}$,
F. Tavecchio$^{6}$
for the MAGIC Collaboration,
S. Buson$^{1}$,
F. D'Ammando$^{7}$
for the \textit{Fermi} Collaboration and
T. Hovatta$^{8}$
for the OVRO Collaboration
}

\afiliations{
$^1$ Delaware University, 19711 Newark, DE, USA \\
$^2$ Universit\`{a} degli Studi di Padova \& INFN Padova, I-35131 Padova, Italy\\
$^3$ Finnish Centre for Astronomy (FINCA) with ESO, University of Turku, Finland\\
$^4$ Tuorla Observatory, University of Turku, Finland\\
$^5$ Universit\`{a} di Siena \& INFN Pisa, I-3100 Siena, Italy\\
$^6$ INAF National Institute for Astrophysics, I-00136 Rome, Italy\\
$^7$ University of Perugia \& INFN, I-06123 Perugia, Italy\\
$^8$ Cahill Center for Astronomy \& Astrophysics, California Institute of Technology, CA 91125, USA\\

}

\email{berger.karsten@gmail.com}

\abstract{We present results of multi-wavelength (MWL) observations of the high-frequency-peaked BL Lacertae (HBL) object 1ES 0806+524 ($z=0.138$, \cite{bade}). Triggered by a high optical state, very high energy (VHE; E $> 100$\,GeV) observations were carried out with the MAGIC stereoscopic system from January to March 2011. During the observations a relatively short VHE $\gamma$-ray flare was detected that lasted no longer than one night. To complement the VHE observations, simultaneous MWL data were collected in high energy $\gamma$-rays using the \textit{Fermi} Large Area Telescope (HE, 300 MeV - 100 GeV), in the X-ray and UV band with the \textit{Swift} satellite, in the optical R--band through observations with the KVA telescope and in the radio band using the OVRO telescope. This constitutes the first time that such a broad band coverage has been obtained for this source. We study the source properties through the characterization of the spectral energy distribution (SED) and its evolution through two different VHE flux states. The SED can be modeled with a simple one-zone SSC model, resulting in parameters that are comparable to those obtained for other HBLs.}


\begin{document}
\maketitle

\section{Introduction}

Active galactic nuclei (AGN) are among the most variable objects in the known universe. Their broad band energy spectrum spanning from radio to VHE $\gamma$-rays can be characterized by two distinct peaks: one in the optical to UV range, commonly interpreted as synchrotron radiation, and a second one in the $\gamma$-ray band that is supposed to originate from inverse Compton scattering of synchrotron photons. The discovery of several AGN in VHE $\gamma$-rays during high optical states \cite{reinthal} leads to the suggestion that optical and VHE $\gamma$-ray high states are connected. In some cases, long--term studies of the broad band emission seem to confirm such a conclusion \cite{aleksic12c}, whereas in other sources this seems to be only partially the case \cite{foschini, aharonian09}.

1ES 0806+524, classified as high peaked BL Lacertae object (based on the peak of the synchrotron emission and a featureless optical spectrum \cite{schachter}) was suggested as a VHE candidate by \cite{costamante}. Several VHE observations have been carried out \cite{horan, aharonian04, aleksic11}, but initially no emission was found. Finally, VERITAS observations from November 2006 to April 2008 succeeded to detect the source in a non-variable, low emission state (integral flux E$>300$\,GeV equal to $\sim$1.8\% of the Crab Nebula flux \cite{acciari09}).

This proceeding reports MAGIC observations during a high optical state in early 2011. We analyze flux and spectral variability in VHE $\gamma$-rays and model the spectral energy distribution including complementary data from contemporaneous observations in HE $\gamma$-rays carried out by the \textit{Fermi} Large Area Telescope (LAT), in X-rays performed by the \textit{Swift} satellite and in the optical R--band by the KVA. Radio data coverage is provided by the OVRO telescope at 15\,GHz.

\section{MAGIC observations}
MAGIC is a stereoscopic system of two 17\,m Imaging Atmospheric Cherenkov Telescopes (IACTs) located at the Roque de Los Muchachos, La Palma, Canary Islands (28.8$^\circ$ N, 17.8$^\circ$ W, 2225\,m a.s.l.). Due to its low energy threshold ($\sim$60\,GeV) and high sensitivity\footnote{E$>290$\,GeV; flux $<0.8$\% of the Crab Nebula in 50\,h \cite{aleksic12d}.}, it is well suited for VHE $\gamma$-ray observations of blazars.

Observations carried out between January and March 2011 on 13 nights for a total of $\sim$24\,h resulted in a flare detection on February 24$^{\mathrm{th}}$ (7.6~$\sigma$ above 250~GeV \cite{mariotti}) and a strong overall detection of $\sim$9.9~$\sigma$ above 250\,GeV within 16.1\,h of effective observation time. The observations were triggered by a high optical state of the source reported by the KVA telescope in the framework of the Tuorla blazar monitoring program.

\section{The MWL context}
\begin{figure*}[t!]
\includegraphics[width=0.95\textwidth]{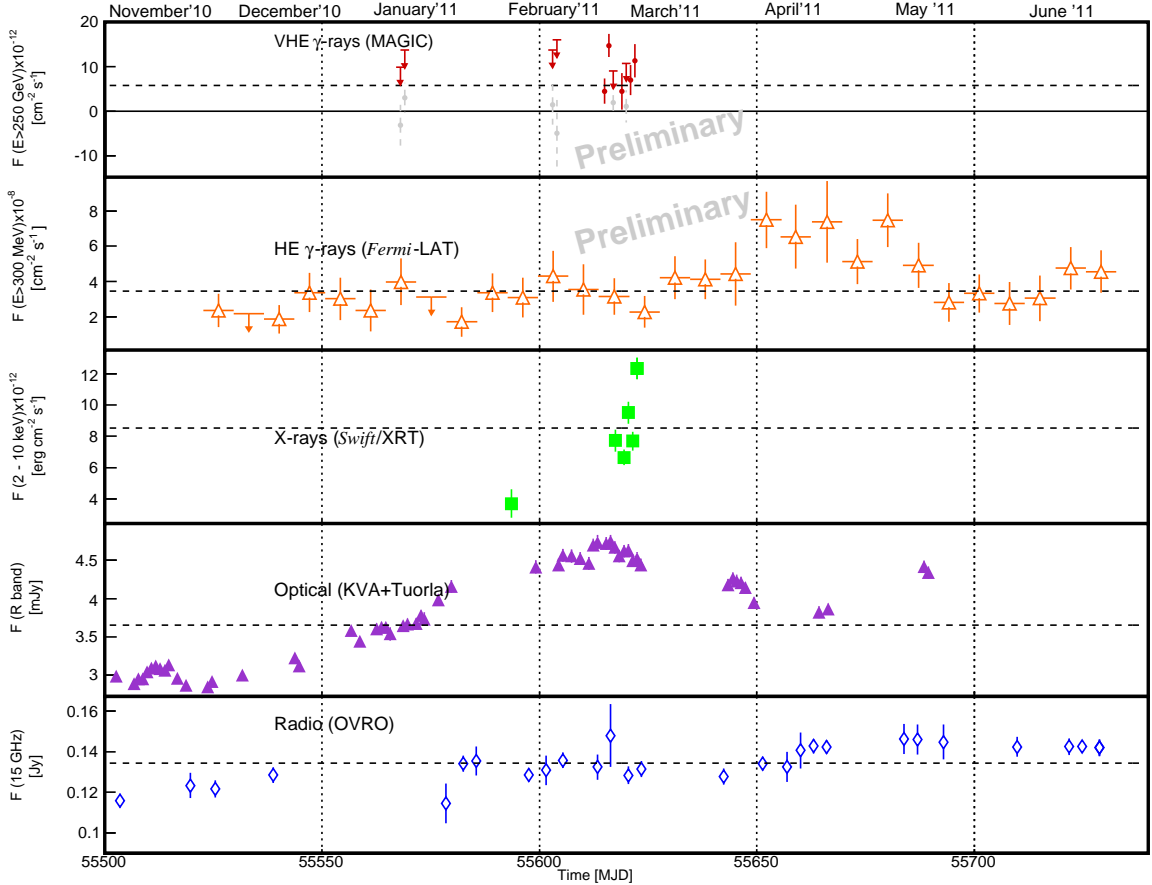}
\caption{From top to bottom: MWL light curve of 1ES 0806+524 from December 2010 to June 2011 covering VHE $\gamma$-rays (red and gray circles) by MAGIC, HE $\gamma$-rays (orange triangles) by \textit{Fermi}--LAT, X-rays (green squares) by \textit{Swift}/XRT, R--band (violet triangles) and radio (blue diamonds) by the KVA and OVRO telescopes. 95\% confidence upper limits are indicated as downward arrows in VHE $\gamma$-rays where the flux is compatible with zero as well as in the HE band for each time bin where the test statistic (TS) value for the source was TS$<$9. The individual light curves are daily binned except for the light curve in the HE band, where a weekly binning has been applied. The mean flux levels of the respective observation periods are indicated (dashed lines).}
\label{fig:LC_MWL}
\end{figure*}

Figure~\ref{fig:LC_MWL} shows the long-term MWL light curves in VHE $\gamma$-rays (MAGIC), HE $\gamma$-rays (\textit{Fermi}-LAT), X-rays (\textit{Swift}), in the R--band (KVA, Tuorla) and radio regime (OVRO telescope).

The integral flux in VHE $\gamma$-rays (E $> 250$\,GeV) of the observation period excluding the flare was estimated to be $(3.1 \pm 0.1_{\mathrm{stat}})\cdot10^{-11}$\,cm$^{-2}$\,s$^{-1}$ corresponding to $(1.9 \pm 0.6)$\% of the Crab Nebula flux, respectively $(2.1 \pm 0.8)$\% C.U. above 300\,GeV. This is in good agreement with the flux level reported by VERITAS during the source detection (1.8\% C.U. above 300 GeV \cite{acciari09}). The integral flux observed during the flare is $(1.4 \pm 0.3_{\mathrm{stat}})\cdot10^{-11}$\,cm$^{-2}$\,s$^{-1}$, which is equal to $(8.6 \pm 1.9)$\% C.U. Since the nights before and after the flaring event showed a significantly lower flux, we assume a short--term variability of the time scale of one day as an upper limit.

The source showed hints of variability in HE $\gamma$-rays evidenced by a low probability of 6\% for constant emission during the period of interest. A smooth flux increase from the beginning of March until mid--April 2011 reaching a maximum of $(7.5 \pm 1.5)\cdot10^{-8}$\,cm$^{-2}$\,s$^{-1}$ (E $>$ 300\,MeV) was observed with a delay compared to the other wavelengths followed by a steady decrease. Given the long integration time of seven days, no clear conclusion regarding simultaneous source variability can be drawn.

In X-rays (2--10\,keV) the time--averaged flux is $(8.7 \pm 1.0)\cdot10^{-12}$\,erg\,cm$^{-2}$\,s$^{-1}$ showing an enhancement on March 2$\mathrm{^{nd}}$ to $(13.0 \pm 0.5)\cdot10^{-12}$\,erg\,cm$^{2}$\,s$^{-1}$, with indications of spectral hardening \cite{stamerra}. Compared to previous observations performed on February 1$^{\rm{st}}$, 2011, the measured flux is 2--3 times higher and is comparable to the flux level measured in March 2008 during the first detection of the source in VHE $\gamma$-rays. Hence, the \textit{Swift}/XRT observations confirm the high activity state of the source exhibiting a clear variability in X-rays with a probability of a constant flux of $6.9\cdot10^{-21}$.

\textit{Swift}/UVOT observations were carried out with different filters in the ultraviolet bands. The brightness of $(14.4 \pm 0.03)$\,mag measured on March $2^{\rm{nd}}$ in the UV-W2 and UV-M2 bands is almost unchanged with respect to February $1^{\rm{st}}$ where a brightness of $(14.5 \pm 0.03)$\,mag was measured. However, 1ES 0806+524 appears about 1\,mag brighter compared to the UV flux observed in March 2008. During the observations the UV band photometry is compatible with a constant flux.

In the R--band, the core flux was seen to have greatly increased over the long--term base level starting from November 2010. Over the following months the flux density continued to increase until reaching a maximum of $(4.75 \pm 0.09)$\,mJy (host galaxy subtracted), an almost threefold increase over the quiescent state of 1.72\,mJy \cite{reinthal} on the night of February 24$^{\mathrm{th}}$, 2011, during the outburst in VHE $\gamma$-rays, after which the flux level began to decrease steadily.

The long--term radio light curve provides a probability of $\sim$6.2$\cdot10^{-5}$ for a non--variable source with an average flux level of 0.14\,Jy. A correlation between the high $\gamma$-ray state and a possible flux increase in the radio band to 0.15\,Jy is observable. However, due to the large statistical error, the flux is compatible with the flux level of the night before and after the VHE $\gamma$-ray flare. Compared to observations from November 2010, the radio data show a marginal flux increase from mid--January to May 2011, exceeding the mean flux level of the overall observation period.

     \begin{table}[h!]
      \centering
    \begin{tabular}{cllcc}
  
  SSC&&\multicolumn{2}{c}{Source state} \\
    parameter&&high&low\\
     \hline
      \hline
      $\gamma _{\rm min}$ &$[10^3 $]&1.0&1.0\\
      $\gamma _{\rm b}$ & [$10^4 $]&2.0&2.0\\
      $\gamma _{\rm max}$&[$10^5 $]  &7.0&7.0\\
       $n_1$&&2.0&2.0\\
       $n_2$&&3.85&3.90\\
       $B$&[G]& 0.05&0.05\\
       $K$ & [$10^3$\,cm$^{-3}]$ &19&4.0\\
     $R$&$[10^{16}$\,cm]& 1.17& 1.90\\
     $\delta $&&28&28\\
  $P_{\rm e}$&[$10^{43}$\,erg/s] &44.8&22.0\\
  $P_{B}$ & [$10^{43}$\,erg/s]&0.10&0.26\\
 $P_{\rm p}$& [$10^{43}$\,erg/s]&7.7&3.3\\
    \end{tabular}
     \caption{Parameters of the SSC model (described in section 4) are reported for the high and low state SED presented in Figure~\ref{fig:SED_high} and~\ref{fig:SED_low}. The derived power carried by electrons, magnetic field and protons is also shown.} 
\label{tab:SED} 
    \end{table}

\section{Theoretical modeling and interpretation}

\begin{figure}[h]
\centering
\includegraphics[width=0.42\textwidth]{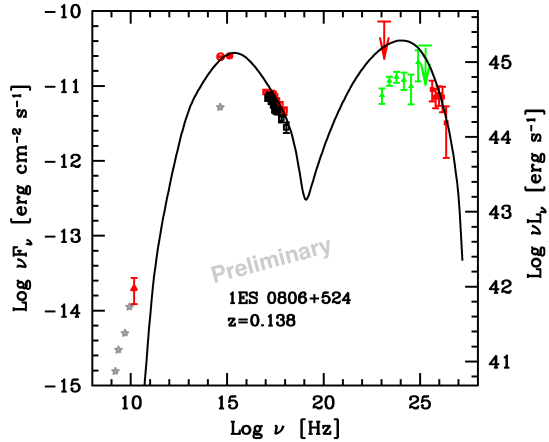}
\caption{SED of the high state activity of 1ES 0806+524 on February $24^{\rm{th}}$. MAGIC VHE $\gamma$-ray data was corrected for EBL absorption \cite{dominguez} (red, blank squares). We show the \textit{Fermi}--LAT upper limit from February $24^{\rm{th}}$ (red arrow) next to an averaged spectrum (green, filled triangles) derived from November 2010 until June 2011. Simultaneous and quasi--simultaneous data from February $27^{\rm{th}}$ and March $2^{\rm{nd}}$ are provided by \textit{Swift}/XRT/UVOT in X-rays (black and red, filled squares) and the U--band (red, filled circles), optical data in the R--band from KVA (red, blank circle) and radio data from the OVRO telescope (red, blank triangle). Archival radio data (gray, blank stars) from NED are also shown. The solid line represents the fit with a one--zone SSC--model (Table~\ref{tab:SED}).}
\label{fig:SED_high}
\end{figure}

\begin{figure}[h]
\centering
\includegraphics[width=0.42\textwidth]{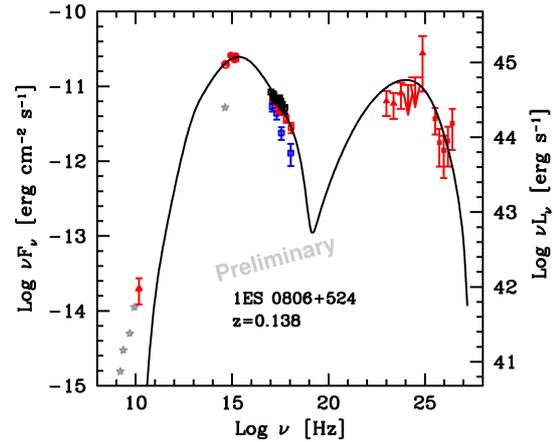}
\caption{Same as Figure~\ref{fig:SED_high}, but for the low state activity of 1ES 0806+524.  The HE $\gamma$-ray spectrum is been derived from \textit{Fermi}--LAT observations carried out from January to March $2^{\rm{nd}}$, 2011, excluding the night of the VHE $\gamma$-ray flare. The simultaneous and quasi--simultaneous data from \textit{Swift}/XRT/UVOT in X-rays is divided as follows: blue, filled squares correspond to observations from February $2^{\rm{nd}}$, while observations from February $28^{\rm{th}}$ and March $1^{\rm{st}}$ are indicated in red and black (due to exact superimposition, data from February $25^{\rm{th}}$ and $27^{\rm{th}}$ are not visible). Red, filled circles in the U--band denote observations from February $2^{\rm{nd}}$ to March $1^{\rm{st}}$.}
\label{fig:SED_low}
\end{figure}

The spectral energy distributions of the source describing the high and low state during the MAGIC observations, together with simultaneous data from \textit{Fermi}, \textit{Swift}, the KVA and OVRO telescopes are presented in Figure~\ref{fig:SED_high} and Figure~\ref{fig:SED_low}. EBL corrections have been applied to the VHE $\gamma$-ray data using \cite{dominguez}. Data from \textit{Swift}/UVOT  and optical data in the R--band provided by the KVA telescope have been corrected for galactic extinction and the host galaxy contribution respectively \cite{fitzpatrick, nilsson}.

Simultaneous and quasi-simultaneous data have been combined in such a way, that they cover the high and low state as observed in VHE $\gamma$-rays. Unfortunately, the \textit{Fermi}--LAT cannot detect 1ES 0806+524 on time scales shorter than one week. To provide simultaneous coverage in this energy band for the high state of the source, an upper limit was calculated for February 24$^{\rm{th}}$ which was included in the SED modeling. In addition, an averaged SED from eight months of observations taken from November 2010 to June 2011 has been derived and is shown for comparison in Figure~\ref{fig:SED_high}. For the low state, an averaged spectrum from January to March 2$^{\rm{nd}}$ was produced in this energy band excluding the night of the $\gamma$-ray flare from the data sample.

Archival data in the radio band (from NED), and the R--band (from \cite{acciari09}) are also shown in gray. The \textit{Swift} data were divided into high state (February 27$^{\rm{th}}$ and March $2^{\rm{nd}}$) and low state observations (February $1^{\rm{st}}$ to March $1^{\rm{st}}$). Since no variability in the radio band was detected during the MAGIC observations, all available data were included for both source states.

MAGIC observations indicate flux variability of a factor of three in VHE $\gamma$-rays, whereas the comparison of the low state data from \textit{Fermi} with the eight--months averaged spectrum does not indicate any significant variation in the HE regime. As observations in the U--band have been carried out in "filter of the day" mode, no explicit comparison between the two activity states of the source can be made. Since a continuous flux increase in the R--band is present, starting well before 1ES 0806+524 showed flaring activity in the VHE regime and in X-rays, that exhibits a plateau with only marginal flux variations towards the end of February, no significant difference is seen in the synchrotron component, while the flux of the inverse Compton peak increases by a factor of 2$\sim$3. The large uncertainty in the HE $\gamma$-ray band limits the determination of the inverse Compton peak.

A one--zone synchrotron--self--Compton (SSC) model is applied to reproduce the SED of both source states (for a detailed description see \cite{mt2003}), where a spherical emission region of radius $R$ is assumed, filled with a tangled magnetic field of intensity $B$.  A population of relativistic electrons is approximated by a smoothed broken power law that is parametrized by the minimum, break and maximum Lorentz factors $\gamma_{\rm{min}}$, $\gamma_{\rm{b}}$ and  $\gamma_{\rm{max}}$ as well as the slopes $n_1$ and $n_2$ before and after the break, respectively. Relativistic effects are taken into account via the Doppler factor $\delta$.

We report the physical parameters derived for both source states in Table~\ref{tab:SED}. The values are similar to those typically inferred for other HBL objects (see e.g. \cite{tavecchio10}). Most of the physical parameters referring to the electron spectrum of the high and low state are compatible, except the electron density $K$, which shows an increase by a factor of $\sim$5 during the high state. Furthermore, the emission region $R$ during the high state is approximately half the radius of the emission region of the low state.

In Table~\ref{tab:SED} we also report the power carried by the jet through electrons ($P_{\rm e}$), magnetic field ($P_{B}$) and protons ($P_{\rm p}$, derived assuming the presence of one cold proton per emitting electron). Magnetic field and protons appear subdominant with electrons carrying most of the jet power, whose total value, $P_{\rm jet}=P_{\rm e}+P_{B}+P_{\rm p}\sim 10^{44}$ erg s$^{-1}$ is also typical (e.g. \cite{ghisellini10}). While the jet power carried by the electrons and protons doubles during the flare, the magnetic field power is reduced. The minimal variability time scale of the model ($\sim$0.3 days based on a causality argument) is compatible with $\sim$one day flux doubling time inferred from the VHE light curve.

\section{Summary \& Discussion}
We reported MAGIC observations of the HBL 1ES 0806+524 in early 2011, triggered by an optical high state. MAGIC caught a short (one day) VHE $\gamma$--ray flare with a VHE flux increase of a factor of three. Excluding the flare, the VHE $\gamma$-ray flux from the source was in good agreement with VERITAS observations in 2008 \cite{acciari09}. 

The MWL data show variability across the electromagnetic spectrum. While the optical and VHE $\gamma$-ray flares have been accompanied by an outburst in X-rays, an increase of the HE flux has been observed later and on longer time scales. Although the VHE observations were optically triggered, no apparent evidence of a correlation was found between these wave bands. 

A one--zone SSC model can explain the spectral energy distributions obtained from the low and high state with parameters similar to those typically inferred for other HBL objects. According to the model, during the high VHE state, the power in electrons and protons increased while the magnetic energy density decreased. The high state SED indicates a mild inverse Compton dominance.

\vspace*{0.5cm}
\footnotesize{{\bf Acknowledgment:}{
We would like to thank the Instituto de Astrof\'{\i}sica de
Canarias for the excellent working conditions at the
Observatorio del Roque de los Muchachos in La Palma.
The support of the German BMBF and MPG, the Italian INFN, 
the Swiss National Fund SNF, and the Spanish MICINN is 
gratefully acknowledged. This work was also supported by 
the Marie Curie program, by the CPAN CSD2007-00042 and MultiDark
CSD2009-00064 projects of the Spanish Consolider-Ingenio 2010
programme, by grant DO02-353 of the Bulgarian NSF, by grant 127740 of 
the Academy of Finland, by the YIP of the Helmholtz Gemeinschaft, 
by the DFG Cluster of Excellence ``Origin and Structure of the 
Universe'', by the DFG Collaborative Research Centers SFB823/C4 and SFB876/C3,
and by the Polish MNiSzW grant 745/N-HESS-MAGIC/2010/0.\\
The $Fermi$ LAT Collaboration acknowledges support from a number of agencies and institutes for both development and the operation of the LAT as well as scientific data analysis. These include NASA and DOE in the United States, CEA/Irfu and IN2P3/CNRS in France, ASI and INFN in Italy, MEXT, KEK, and JAXA in Japan, and the K.~A.~Wallenberg Foundation, the Swedish Research Council and the National Space Board in Sweden. Additional support from INAF in Italy and CNES in France for science analycsis during the operations phase is also gratefully acknowledged.\\
The OVRO 40-m monitoring program is
supported in part by NASA grants NNX08AW31G 
and NNX11A043G, and NSF grants AST-0808050 
and AST-1109911.}}


\begin{thebibliography}{}

\bibitem{bade}N. Bade, et al., A\&A 334 (1998), 459-472. 

\bibitem{reinthal}R. Reinthal, et al., J. Phys.: Conf. Ser. 355 (2012), 012013 doi:10.1088/1742-6596/355/1/012013.

\bibitem{aleksic12c}J. Aleksi\'c, et al. (MAGIC Coll.), Astrophys. J. 748 (2012), 46 doi:10.1088/0004-637X/748/1/46.

\bibitem{foschini}L. Foschini,  et al., ApJ 657 (2007), L81-L84 doi:10.1086/513271. 

\bibitem{aharonian09}F. Aharonian, et al., ApJ 696 (2009), L150-L155 doi:10.1088/0004-637X/696/2/L150.

\bibitem{schachter}J. Schachter, et al., ApJ 412 (1993), 541-549 doi:10.1086/172942.

\bibitem{costamante}L. Costamante and G. Ghisellini, Astron. Astrophys 384 (2002), 56-71 doi:10.1051/0004-6361:20011749.

\bibitem{horan}D. Horan, et al. (Whipple Coll.), ApJ 603 (2004), 51-61 doi:10.1086/381430.

\bibitem{aharonian04}F. Aharonian, et al. (HEGRA Coll.), A\&A 421 (2004), 529.

\bibitem{aleksic11}J. Aleksi\'c, et al. (MAGIC Coll.), ApJ 729 (2011), 115-125 doi:10.1088/0004-637X/729/2/115.

\bibitem{acciari09}V. Acciari, et al. (VERITAS Coll.), ApJL 690 (2009), L126-L129 doi:10.1088/0004-637X/690/2/L126.

\bibitem{aleksic12d}J. Aleksi\'c, et al. (MAGIC Coll.), Astroparticle Physics 35 (2012), 435-448 doi:10.1016/j.astropartphys.2011.11.007. 

\bibitem{mariotti}M. Mariotti, et al. (MAGIC Coll.), ATel 3192 (2011).

\bibitem{stamerra}A. Stamerra, et al., ATel 3208 (2011).

\bibitem{fitzpatrick}E.~L. Fitzpatrick, PASP 111 (1999), 63-75 doi:10.1086/316293.

\bibitem{nilsson}K. Nilsson, et al., A\&A 475 (2007), 199-207 doi:10.1051/0004-6361:20077624.

\bibitem{dominguez}A. Dom\'inguez, et al., MNRAS 410 (2011), 2556-2578.

\bibitem{mt2003}L. Maraschi and F. Tavecchio, ApJ 593 (2003), 667-675 doi:10.1086/342118.

\bibitem{tavecchio10}F. Tavecchio, et al., MNRAS 401 (2010), 1570-1586.

\bibitem{ghisellini10}G. Ghisellini, et al., MNRAS 402 (2010), 497-517 doi:10.1111/j.1365-2966.2009.15898.x.


\end{thebibliography}
\end{document}